\documentstyle[12pt]{article}
\def\fracl#1#2{\frac{\displaystyle #1}{\displaystyle #2}}

\begin{document}

\begin{center}
{\sc V.Bashkov, M.Malakhaltsev}\\
{\large\bf Relativistic mechanics on rotating disk
}
\end{center}

In this paper we find equations of particle motion from the point of view of 
observer
 on a rotating disk.
In [1] we found a local transformation from the inertial reference system to 
the noninertial rotating reference system, which generalizes the Lorentz
transfromation. This transformation can be written in  the following form with 
respect to cylindrical coordinates:
$$
dr' = dr, \quad 
r'd\phi' = \fracl{r' d\phi- \fracl{\omega r dt}{c^2}}{\sqrt{1- 
\fracl{\omega^2 r^2}{c^2}}}, 
\quad
dt' = \fracl{dt - \fracl{\omega r}{c^2} r d\phi}{
\sqrt{1 - \fracl{\omega^2 r^2}{c^2}}}
\eqno{(1)}
$$
   
Moreover, in [1] we have proved that the transformation (1) 
preserves the Lorentz metric, 
$$
ds^2 = -dr^2 - r^2 d\phi^2 + c^2 dt^2 = 
-{dr'}^2 - {r'}^2 d{\phi'}^2 + c^2 d{t'}^2.
\eqno{(2)}
$$
The transformation (1) is nonholonomic, i.e. the 
transformation matrix 
$$
||A^{i'}_i || =
\left(
\begin{array}{ccc} 
1 & 0 & 0 \\ 
0 &  \fracl{1}{\sqrt{1-\fracl{\omega^2 r^2}{c^2}}}  & 
\fracl{-\omega}{\sqrt{1-\fracl{\omega^2 r^2}{c^2}}} \\ 
0 &  \fracl{-\omega r^2}{c^2\sqrt{1-\fracl{\omega^2 r^2}{c^2}}}  & 
\fracl{1}{\sqrt{1-\fracl{\omega^2 r^2}{c^2}}} \\ 
\end{array}
\right)
$$
is not a Jacobian matrix of any coordinate transformation.
This is important when we are constructing mechanics on rotaing disk.
We suppose that with respect to the noninertial reference system the particle 
motion is subordinate to the the law
$$
\fracl{\delta v^{i'}}{\delta s} \equiv \fracl{dv^{i'}}{ds'} 
+ \Gamma^{i'}_{j'k'}\fracl{dx^{j'}}{ds'} \fracl{dx^{k}}{ds'}
=F^{i'}/m_0 \eqno{(3)},
$$
where $m_0$ is the rest mass of particle, 
$v^{i'}\equiv \fracl{dx^{i'}}{ds}$,
$F^{i'}$ is the effective force in the noninertial reference system,
$\Gamma^{i'}_{j'k'}$ are the  connection coefficients with respect to
 the noninertial reference system, which are obtained from the Christoffel 
symbols $\Gamma^i{i}_{jk}$ of the metric (2) in the following way: 
$$
\Gamma^{i'}_{j'k'} = A^{i'}_k A^j _{j'} \fracl{\partial A^k_{k'}}{\partial x^j}
+ \Gamma^{i}_{jk}  A^{i'}_i  A^k_{k'} A^j_{j'}.
\eqno{(4)}
$$
After easy calculations we get
$$
\begin{array}{l}
\Gamma^{0'}_{0'1'} = -\fracl{\omega^2 r/c^2}{ 1-\fracl{\omega^2 r^2}{c^2}};
\quad
\Gamma^{0'}_{1'2'} = -\Gamma^{0'}_{2'1'} =
\fracl{\omega r/c^2}{(1-\fracl{\omega^2 r^2}{c^2})};
\\
\Gamma^{1'}_{0'0'} =
 -\fracl{r \omega^2}{1-\fracl{\omega^2 r^2}{c^2}};
\quad
\Gamma^{1'}_{0'2'} =\Gamma^{1'}_{2'0'} =
 -\fracl{r \omega}{1-\fracl{\omega^2 r^2}{c^2}};
\quad
\Gamma^{1'}_{2'2'} = -\fracl{r}{1-\fracl{\omega^2 r^2}{c^2}}; 
\\
\Gamma^{2'}_{0'1'} = \Gamma^{2'}_{1'0'} =
\fracl{\omega}{r(1-\fracl{r\omega^2 r^2}{c^2})};
\quad
\Gamma^{2'}_{1'2'} = \fracl{1}{r};
\quad
 \Gamma^{2'}_{2'1'} = 
\fracl{1}{r(1-\fracl{\omega^2 r^2)}{c^2}};
\end{array}
\eqno{(5)}
$$
The other connection coefficients are zero.

Now let us write the motion equation for a particle with nonzero rest 
mass ($ds^2 \ne 0$) and for a particle with zero rest mass
($ds^2 = 0$) , with respect to 
the noninertial reference system on rotating disk.

Using (5), we get for the case $ds^2 \ne 0$

$$
\begin{array}{l}
\fracl{d^2 t}{ds^2} - \fracl{\omega^2 r / c^2}{1-\fracl{\omega^2 r^2}{c^2}}
\fracl{dt}{ds} \fracl{dr}{ds} = \fracl{1}{m} F^0;
\\
\fracl{d^2 r}{ds^2} - \fracl{r \omega^2 }{
1-\fracl{\omega^2 r^2}{c^2}}(\fracl{dt}{ds})^2
- \fracl{2 \omega r}{1-\fracl{\omega^2 r^2}{c^2}} 
\fracl{dt}{ds}\fracl{d\phi}{ds} -
\fracl{r}{1-\fracl{\omega^2 r^2}{c^2}} (\fracl{d\phi}{ds})^2
 = \fracl{1}{m} F^1;
\\
\fracl{d^2\phi}{ds^2} =  \fracl{2\omega}{r(1-\fracl{\omega^2 r^2}{c^2})}
\fracl{dt}{ds} \fracl{dr}{ds}
+ \fracl{2-\fracl{\omega^2 r^2}{c^2}}{r(1-\fracl{\omega^2 r^2}{c^2})}
\fracl{dr}{ds} \fracl{d\phi}{ds} =
\fracl{1}{m} F^2
\end{array} 
$$
For a photon ($ds^2 =0$), we set $\frac{dx^i}{d\sigma} = K^i$,
where $\sigma$ is the canonical parameter, and 
$K^i$ are components of the wave vector.
We get

$$
\begin{array}{l}
\fracl{dK^0}{d\sigma} - \fracl{\omega^2 r/c^2}{1-\fracl{\omega^2 r^2}{c^2}} 
 K^0  K^1 = F^0;
\\
\fracl{d K^1}{d\sigma} - \fracl{\omega^2 r}{
1-\fracl{\omega^2 r^2}{c^2}} (K^0)^2 
- \fracl{2 \omega r}{1-\fracl{\omega^2 r^2}{c^2}} 
K^0 K^2 -  \fracl{r}{1-\fracl{\omega^2 r^2}{c^2}} (K^2)^2 = F^1;
\\
\fracl{d K^2}{d\sigma} =  \fracl{2\omega}{r(1-\fracl{\omega^2 r^2}{c^2}}
 K^0 K^1 
+ \fracl{2-\fracl{\omega^2 r^2}{c^2}}{r(1-\fracl{\omega^2 r^2}{c^2})}
   K^1 K^2 =
 F^2
\end{array} 
$$

Thus we see that a particle moving along a rotating disk is influenced by 
forces arising from geometry, besides the external forces $F^i$. These forces 
connected with nonholonomity of reference system are  generated by 
the noninertiality. They  can be considered as analogs of the centrifugal forces and
the Coriolis forces.

\begin{center}
References
\end{center}

1. V.Bashkov, M. Malakhaltsev {\em Geometry of rotating disk and Sagnac effect},
arXiv: gr-qc/0011061, 18 Nov. 2000.
\bigskip

{\bf Address:} {\it Kazan State University, Kremlevskaya, 18, Kazan: 420008, Russia}

{\bf E-mail address}: {\it Victor.Bashkov@ksu.ru, Mikhail.Malakhaltsev@ksu.ru}

\end{document}